\documentclass{article}

\usepackage{spconf,amsmath,graphicx}
\usepackage{xcolor,soul,framed} 

\colorlet{shadecolor}{yellow}

\usepackage{array}
\usepackage{url}

\usepackage{cite}
\usepackage{booktabs} 
\usepackage{multirow}
\usepackage{multicol}
\usepackage{amsfonts}
\usepackage{float}
\usepackage{tabularx}

\usepackage{amssymb}
\usepackage{pifont}
\newcommand{\cmark}{\ding{51}}%
\newcommand{\xmark}{\ding{55}}%

\title{ImagineNET: Target Speaker Extraction with Intermittent Visual Cue \\ through Embedding Inpainting}
\name{Zexu Pan~$^{1,2}$, Wupeng Wang~$^{2}$, Marvin Borsdorf~$^{3}$,  Haizhou Li~$^{4,2,3}$
\thanks{This work is supported by 1) German Research Foundation (DFG) under Germany's Excellence Strategy (University Allowance, EXC 2077, University of Bremen); 2) Huawei Noah's Ark Lab; 3) National Natural Science Foundation of China (Grant No. 62271432), Internal Project Fund from Shenzhen Research Institute of Big Data (Grant No. T00120220002); 4) Guangdong Provincial Key Laboratory of Big Data Computing, The Chinese University of Hong Kong, Shenzhen (Grant No. B10120210117-KP02); 5) Agency of Science, Technology and Research (A*STAR) under its AME Programmatic Funding Scheme (Project No. A18A2b0046).}
}
\address{
  $^1$Institute of Data Science, National University of Singapore (NUS), Singapore\\
  $^{2}$Department of Electrical and Computer Engineering, NUS, Singapore\\
  $^{3}$Machine Listening Lab, University of Bremen, Germany\\
  $^{4}$Shenzhen Research Institute of Big Data, School of Data Science,\\ The Chinese University of Hong Kong, Shenzhen, China
  }
\begin{document}
\ninept
\maketitle
\setlength{\abovedisplayskip}{4pt}
\setlength{\belowdisplayskip}{4pt}

\begin{abstract}
The speaker extraction technique seeks to single out the voice of a target speaker from the interfering voices in a speech mixture. Typically an auxiliary reference of the target speaker is used to form voluntary attention. Either a pre-recorded utterance or a synchronized lip movement in a video clip can serve as the auxiliary reference. The use of visual cue is not only feasible, but also effective due to its noise robustness, and becoming popular. However, it is difficult to guarantee that such parallel visual cue is always available in real-world applications where visual occlusion or intermittent communication can occur. In this paper, we study the audio-visual speaker extraction algorithms with intermittent visual cue. We propose a joint speaker extraction and visual embedding inpainting framework to explore the mutual benefits. To encourage the interaction between the two tasks, they are performed alternately with an interlacing structure and optimized jointly. We also propose two types of visual inpainting losses and study our proposed method with two types of popularly used visual embeddings. The experimental results show that we outperform the baseline in terms of signal quality, perceptual quality, and intelligibility.
\end{abstract}
\begin{keywords}
Cocktail party problem, multi-modal, target speaker extraction, visual occlusions, inpainting
\end{keywords}
%

\section{Introduction}
\label{sec:introduction}
\vspace{-.1cm}

Speech is the most natural way of communication between humans. Therefore, the study and development of human-machine interaction systems, such as active speaker detection~\cite{tao2021someone}, speaker localization~\cite{qian2021multi}, speech recognition~\cite{wang2022predict}, lyrics transcription~\cite{gao2021tran}, and emotion recognition~\cite{pan2020multi} constitutes an important part in today's research. However, these algorithms are adversely affected by the presence of interference speakers and acoustic noise. Speech separation, as well as speaker extraction, are techniques to solve those problems.

With speech separation~\cite{luo2019conv,hershey2016deep,von2022sasdr,wang2023time}, we seek to disentangle the overlapping voices in a given speech mixture into the individual speaker voices in one step. Speaker extraction~\cite{wang2019voicefilter,vzmolikova2019speakerbeam,Chenglin2020spex,ephrat2018looking,wu2019time,pan2021reentry,usev21,afouras2018conversation,Li2020,tavcse2022,pan2020muse,marvin2021,pan2022hybrid} takes a different approach and extracts only the voice of a single target speaker. This method relies on an auxiliary reference that directs the system's attention towards the target. Examples of such auxiliary references are pre-recorded reference speech~\cite{vzmolikova2019speakerbeam,Chenglin2020spex,wang2019voicefilter}, speech-synchronized video clips~\cite{ephrat2018looking,wu2019time,usev21,pan2021reentry,afouras2018conversation}, or content information~\cite{Li2020,tavcse2022}. 

Among the auxiliary references for speaker extraction algorithms, lip recording has been shown to be the most direct and informative due to its noise robustness. The TDSE model~\cite{wu2019time} applies a pre-trained visual speech recognition (VSR) network as a visual encoder to extract viseme-phoneme mapping embeddings that can be used as a cue for the target speaker. The \textit{reentry} model~\cite{pan2021reentry} analyzes the visual-only or audio-visual speech-lip synchronization embeddings as the cue through a self-supervised pre-trained speech-lip synchronization detection network (SLSyn).

\begin{figure*}[th]
  \centering
  \includegraphics[width=\linewidth]{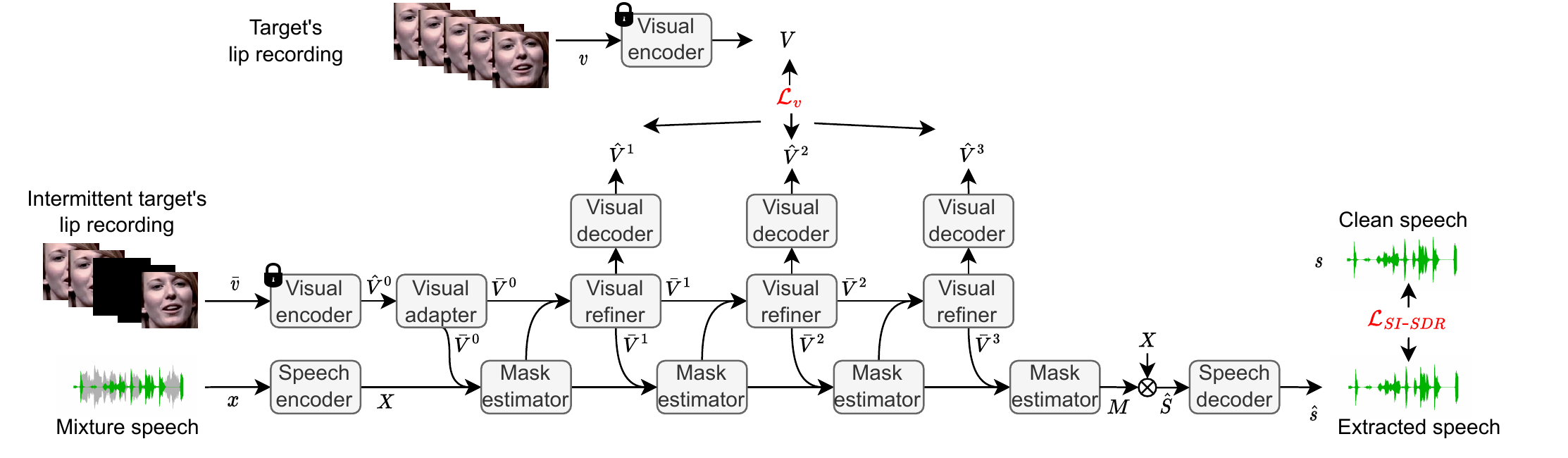}
  \vspace*{-8mm}
  \caption{Proposed joint speaker extraction and visual embedding inpainting network (ImagineNET). It consists of a speech encoder, a visual encoder, a visual adapter, $R$ repeated mask estimators, $R-1$ repeated visual refiners and visual decoders, and a speech decoder. We illustrate the model with $R=4$ here. The symbol $\otimes$ refers to element-wise multiplication. The lock sign on the visual encoder indicates that its weights have been learned on a pretext task and are fixed during the training of the speaker extraction part. When two arrows merge, we apply the channel-wise frame-level concatenation method as our fusion strategy.}
  \vspace*{-4mm}
\label{fig:network}
\end{figure*}

In prior studies, lip movement was commonly expected to be visible throughout the entire session. However, the target speaker's frontal view may not be perfectly visible all the time in everyday conversational situations due to body movement, visual occlusion, or communication interruption. In such cases, the visual reference signal is intermittent. Consequently, this adversely affects those speaker extraction algorithms which rely on visual cues as the reference. In this work, we aim to address the issue of intermittent visual reference in audio-visual speaker extraction.

Related speech enhancement studies~\cite{9053730,9414097} propose to switch from an audio-visual variational auto-encoder (VAE) to an audio-only VAE if the visual reference is noisy or missing. For speaker extraction, the VS model~\cite{afouras2019my} learns a speaker embedding during the first pass, when the lip recording is available, and relies only on the speaker embedding to extract the target speaker's voice during the second pass, when the lip recording is missing. A recent work~\cite{wu2022occl} uses audio features to select relevant visual features in combination with an attention mechanism and a data augmentation strategy in order to address low-resolution, lip-occlusion, and out-of-sync problems altogether. The SEG network~\cite{pan2022seg} avoids the need for lip recordings since it utilizes the co-speech gesture cue from the upper-body video recording as a reference, which is less prone to occlusions.

Instead of discarding the visual reference when the signal is intermittent, the visual inpainting research~\cite{kim2022visualimpainting,liu2021fuseformer} offers an alternative approach that allows the reconstruction of a corrupted video based on audio-visual correspondence. The inpainting idea motivates us to design a joint speaker extraction and visual embedding inpainting network, named ImagineNET, that makes use of audio-visual complementary information. To encourage the interaction between the audio and visual tasks, they are performed alternately and repeatedly with an interlacing structure. In this way, the estimated intermediate speech contributes to the visual embedding inpainting, while the inpainted visual embedding, in turn, helps to refine the speaker extraction mask. The two tasks are jointly optimized. We study our method with two visual embeddings from VSR pre-trained network~\cite{wu2019time} or SLSyn pre-trained network~\cite{pan2021reentry} respectively. We also analyze two ways of learning visual inpainting, namely reconstruction with mean squared error loss and contrastive predictive coding with infoNCE loss~\cite{oord2018representation}. The experimental results show that we outperform our baselines in terms of signal quality, perceptual quality, and speech intelligibility.

\vspace{-.2cm}
\section{Joint speaker extraction and embedding inpainting network}
\label{sec:methodology}
\vspace{-.1cm}

\subsection{Fundamentals of speaker extraction with visual cue}
\vspace{-.1cm}

Let $x$ be a speech mixture signal that consists of the target speech signal $s$ and other interference speech signals. The task of target speaker extraction with visual cue is to estimate $\hat{s}$, which approximates $s$, under the guidance of the target speaker's lip recording $v$.

A time-domain speaker extraction network that utilizes visual cues, such as TDSE~\cite{wu2019time} and USEV~\cite{usev21}, usually consists of five parts: 1) A one-layer convolutional neural network (CNN), called speech encoder, that transforms the raw input mixture waveform $x$ into a frame-based embedding $X$. 2) A visual encoder that extracts the visual embedding $V$ from raw visual input frames $v$. 3) A visual adapter that adapts $V$ towards the speaker extraction task. 4) $R$ repeated mask estimators which refine a mask $M$ that only lets the target speaker's signal pass in $X$. 5) A speech decoder that transforms the masked speech $\hat{S}$ back into a waveform $\hat{s}$.

\vspace{-.2cm}
\subsection{Visual embedding inpainting}
\vspace{-.1cm}

State-of-the-art audio-visual speaker extraction algorithms adopt transfer learning methods for the visual encoder part. Commonly, the visual encoder is pre-trained on other tasks, such as face recognition~\cite{ephrat2018looking}, VSR~\cite{wu2019time}, or speech-lip synchronization detection~\cite{pan2021reentry}. In this way, the knowledge that has been learned from the pretext task~\cite{wu2019time} can be re-used and potentially leverage abundant unlabeled in-domain data with self-supervised learning~\cite{pan2021reentry}. The weights of the visual encoder are usually fixed during the training of the speaker extraction part. Consequently, we prefer studying the inpainting of the visual embeddings from the pre-trained encoder over the inpainting of raw visual input frames, such that our framework can be generalized to any pre-trained visual encoder.

Given an intermittent lip recording $\bar{v}$, missing or occluded images within the stream are set to zero signals before passing them into the pre-trained visual encoder. The task of visual embedding inpainting aims to estimate a $\hat{V}$ that approximates $V$, in which the latter is the output of the pre-trained visual encoder with the non-intermittent target's lip recording as the input.

\vspace{-.2cm}
\subsection{ImagineNET architecture}
\vspace{-.1cm}

Our work is inspired by the success of the \textit{reentry} model~\cite{pan2021reentry}, which shows that alternately refinement of the target speaker mask $M$ and target speaker self-enrollment with an interlacing structure boost the performance of each other. 

We propose a joint speaker extraction and visual embedding inpainting network, referred to as ImagineNET, which is shown in Fig.~\ref{fig:network}. We extend the TDSE network~\cite{wu2019time,pan2021reentry} to deal with the intermittent lip recording $\bar{v}$. More precisely, we introduce an ensemble of $R-1$ visual refiners to gradually refine the visual embedding $\bar{V}^{r-1}$, where $r \in [1, ... , R-1]$. The interaction between the visual embedding inpainting and speaker extraction task relies on an interlacing structure between the visual refiners and the mask estimators, in which the output of one is connected to the input of the other, to analyze the complementary information. The output of each visual refiner is passed through a visual decoder to obtain the inpainted visual embedding $\hat{V}$ in different refinement steps. The visual refiner and visual decoder process every frame of the visual embedding, regardless of whether it is missing in $\bar{v}$ or not, so that our model does not need the prior knowledge of which frames are missing. Another reason is that the embeddings of the non-missing frames could be adversely affected by the missing frames due to the use of kernel and stride on the time dimension in the pre-trained visual encoder.

To simplify matters, Fig.~\ref{fig:network} does not show that the output of each mask estimator is element-wise multiplied with $X$ and passes through both the speech decoder and speech encoder before feeding it into the visual refiner. We share the network's weights for the speech encoders and speech decoders respectively, but not among the visual decoders, visual refiners, and mask estimators. 

We examine our framework on two visual embeddings. The first one is the viseme-phoneme embedding $vsr_v$ that is extracted by the visual front-end of a VSR network~\cite{wu2019time}. The second one is the video-only SLSyn embedding $sync_v$ which is extracted by the visual front-end of the SLSyn network and trained on speech-lip synchronization detection~\cite{pan2021reentry}. 

\vspace{-.2cm}
\subsection{Multi-task learning}
\vspace{-.1cm}

We propose to jointly optimize the speech extraction and the visual inpainting, to foster the network to learn the interaction between those two tasks. The overall loss $\mathcal{L}_{total}$ consists of two parts. The first one is given by the scale-invariant signal-to-distortion ratio \mbox{(SI-SDR)~\cite{le2019sdr}} $\mathcal{L}_{SI\mbox{-}SDR}$ between the extracted speech $\hat{s}$ and the ground truth clean speech $s$. The second one is given by a visual inpainting loss $\mathcal{L}_{v}$ between the ground-truth non-missing visual embedding $V$ and every visual decoder output $\hat{V}^r$, with $r \in [1, ... , R-1]$. 
\begin{equation}
    \mathcal{L}_{total} = \mathcal{L}_{SI\mbox{-}SDR} (s, \hat{s}) + \gamma * \sum_{r=1}^{R-1}\mathcal{L}_{v} (V, \hat{V}^r)
    \label{eqa:overall}
\end{equation}
where
\begin{equation}
    \label{eqa:loss_sisnr}
    \mathcal{L}_{SI\mbox{-}SDR} (s, \hat{s}) = - 10 \log_{10} ( \frac{||\frac{<\hat{s},s>s}{||s||^2}||^2}{||\hat{s} - \frac{<\hat{s},s>s}{||s||^2}||^2})
\end{equation}
With $\gamma$, we adjust the impact of the visual inpainting loss. We examine two ways of learning visual embedding inpainting, each associated with a different loss function as $\mathcal{L}_{v} \in  [\mathcal{L}_{MSE}$, $\mathcal{L}_{infoNCE}]$.

\vspace{-.2cm}
\subsubsection{Reconstruction}
\vspace{-.1cm}

Both mean absolute error (MAE) and mean squared error (MSE) losses have been widely used to reconstruct audio signals~\cite{morrone2021audio,8867915} or images~\cite{kingma2013auto,elharrouss2020image,liu2021fuseformer}. In our work, we consider the MSE loss to measure the reconstruction quality of missing visual embeddings during training:
\begin{equation}
    \label{eqa:loss_mse}
    \mathcal{L}_{MSE} (V, \hat{V}) = \frac{1}{T} \sum_{t=1}^{T} (V_t - \hat{V}_t )^2
\end{equation}
where $t \in [1, ..., T]$ and $T$ is the total number of visual embedding frames.

\vspace{-.2cm}
\subsubsection{Contrastive predictive coding}
\vspace{-.1cm}

Contrastive predictive coding (CPC)~\cite{oord2018representation,he2020momentum} is widely used in representation learning to extract embeddings. It applies categorical cross-entropy to identify the positive sample among a set of noise samples, i.e., infoNCE loss~\cite{oord2018representation}. In our experimental design, the positive sample for a missing visual embedding is given by its corresponding ground-truth embedding, while the negative or rather noise samples are the non-corresponding ground-truth embeddings.

While speaking, our lips have a very high dynamic, which makes the reconstruction challenging when the video stream is missing for some seconds. CPC takes a different approach compared to reconstruction, and maximizes the mutual information between the corresponding decoded visual embeddings and the ground-truth visual embeddings. The loss function is formulated as follows:

\begin{equation}
    \label{eqa:loss_infonce}
    \mathcal{L}_{infoNCE} (V, \hat{V}) = \sum_{i=1}^T -log \frac{exp(\hat{V}_i \cdot V_i / \kappa)}{\sum_{j=1}^T exp(\hat{V}_i \cdot V_j / \kappa)} 
\end{equation}
where $\kappa$ is the temperature parameter and set to 0.07, following the paper~\cite{he2020momentum}.

\newcolumntype{Y}{>{\centering\arraybackslash}X}
\begin{table*}
    \centering
    \caption{Comparison of our proposed ImagineNET with the baseline TDSE under conditions with (\cmark) or without (\xmark) partially missing visual reference during training and inference. Different visual embeddings and visual inpainting losses are used.}
    \addtolength{\tabcolsep}{-3.5pt}
    \resizebox{0.95\linewidth}{!}{
    \begin{tabularx}{\textwidth}{@{}Y|Y|Y|Y|Y|Y|Y|Y|Y@{}} 
       \toprule
        \mbox{System (Sys.)} & Model  &V missing   &\mbox{V embedding} &V loss &SI-SDR &SDR  &PESQ  &STOI\\
        \midrule    
        0   &Mixture    &-   &-   &-   & \ -0.09  & \ \ 0.00   &1.886    &0.632 \\
        
        \midrule
        1   &TDSE &\multirow{3}{*}{\cmark} &\multirow{3}{*}{$vsr_v$} &- & \ \ 9.95 &10.48  &2.809  &0.840\\
        2   &ImagineNET &&    &MSE    &10.81 &11.33   &2.894  &0.854 \\
        3   &ImagineNET &&    &InfoNCE    &10.89	&11.40	&2.903	&0.855 \\
        \midrule
        4   &TDSE &\multirow{3}{*}{\cmark} &\multirow{3}{*}{$sync_v$} &- & \ \ 9.29 & \ \ 9.87  &2.762  &0.825  \\
        5   &ImagineNET &&   &MSE    & \ \ 9.84 &10.46    &2.819  &0.833 \\
        6   &ImagineNET &&   &InfoNCE    & \ \ 9.81	&10.42	&2.824	&0.833 \\
        \midrule
        7   &TDSE &\multirow{3}{*}{\xmark} &\multirow{3}{*}{$vsr_v$} &- &11.88 &12.30 &2.991  &0.878 \\
        8   &ImagineNET &&    &MSE    &12.39 &12.77    &3.023    &0.885 \\
        9   &ImagineNET &&    &InfoNCE &11.99   &12.38  &2.988  &0.881\\
       \bottomrule
    \end{tabularx}
    }
    \addtolength{\tabcolsep}{3.5pt}
    \vspace*{-4mm}
    \label{tab:baseline}
\end{table*}

\vspace{-.2cm}
\section{Experimental Setup}
\label{sec:experiment}
\vspace{-.1cm}

\subsection{Dataset}
\vspace{-.1cm}

In our experiments, we follow~\cite{pan2020muse,pan2021reentry} and use VoxCeleb2-2mix, which is a two-speaker mixtures dataset simulated from the VoxCeleb2~\cite{Chung18b} dataset. It consists of 20,000, 5,000, and 3,000 speech mixture utterances for train, validation, and testing respectively. The target speech is mixed with a random interference speech at a random signal-to-noise ratio (SNR) in a range from -10 dB to 10 dB. The audio sampling rate is 16 kHz and the face track video of the target speaker is provided at 25 frames-per-second (FPS).

We simulate the intermittent visual cue by blacking out the visual signal, starting at a random start frame for a random duration, while keeping the audio signal intact. To put the system on a stress test, the percentage of missing visual frames varies from 0\% to 100\%. This simulates the scenarios in which the face-tracking algorithm fails to detect the presence of the target speaker due to various reasons, such as occlusion or signal interruption.

\vspace{-.2cm}
\subsection{Baseline}
\vspace{-.1cm}

We use the TDSE architecture~\cite{wu2019time} for our baseline, which represents the recent advances in utilizing visual cue for speaker extraction. Compared with our proposed ImagineNET, TDSE does not have visual refiner modules, visual decoder modules, and visual inpainting loss. The original implementation of the TDSE network utilizes $vsr_v$ with a pre-trained VSR network as visual encoder. We implement two TDSE networks with $vsr_v$ and $sync_v$ visual embeddings respectively, in order to form our baselines.

\vspace{-.2cm}
\subsection{Implementation details}
\vspace{-.1cm}

The network architectures of the speech encoder, mask estimators, and speech decoder follow the original TDSE network~\cite{wu2019time}, while the visual adapter follows the implementation in the \textit{reentry} model. The visual refiner adopts the TCN structure similar to the mask estimator but with a smaller dilation. The visual decoder consists of several CNN layers~\footnote{The data generation and the ImagineNET model training codes are available at: \url{https://github.com/zexupan/ImagineNET}}.

Following the original symbol definition for hyperparameters in~\cite{luo2019conv,wu2019time,pan2021reentry}, we set $N, L, B, H, P, X, R$ to 256, 40, 256, 512, 3, 7, and 4 respectively to initialize the ImagineNET. For the TDSE network, we set the hyperparameters in the exact same manner as for ImagineNET, except that we increase $B$ to 384 to raise the model capacity. The total numbers of network parameters are comparable between the ImagineNET (15.8 million) and the TDSE network (16.0 million) if the $vsr_v$ embedding is used.

We empirically adjust $\gamma$ using the validation dataset with the VSR embedding. We find $\gamma = 1$ to achieve the best SI-SDR result. During training, we use Adam optimizer with an initial learning rate of $1e^{-3}$. The learning rate is halved if the best validation loss (BVL) does not improve within six consecutive epochs, and the training stops if the BVL does not improve within ten consecutive epochs. 

\begin{figure}[t]
\begin{minipage}[t]{.485\linewidth}
  \centering
  \centerline{\includegraphics[width=\linewidth]{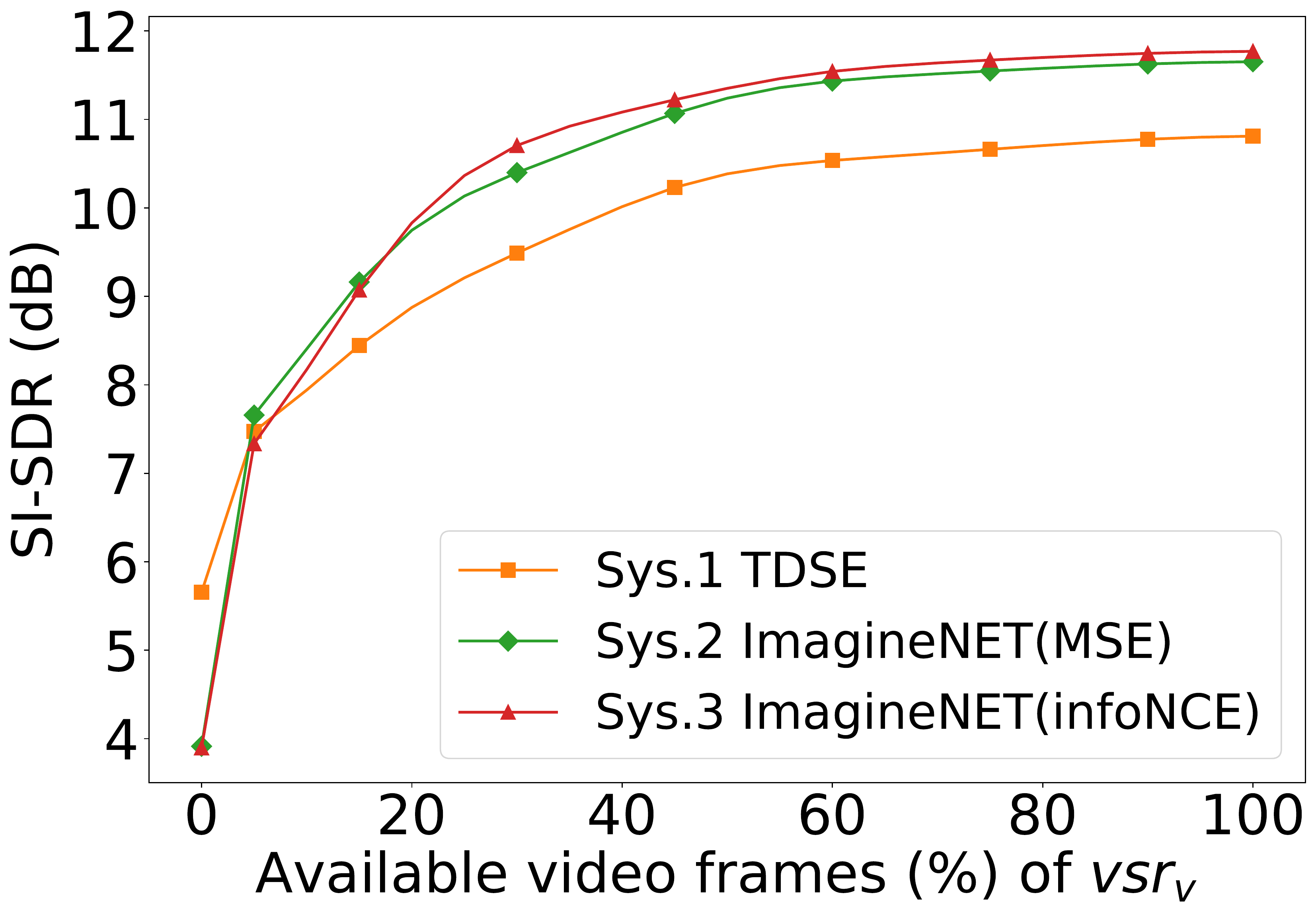}}
  \vspace*{-2mm}
  \caption{\footnotesize The SI-SDR for different percentages of available video frames when $vsr_v$ embedding is used.}\medskip
  \label{fig:occl_vsr}
\end{minipage}
\hfill
\begin{minipage}[t]{.485\linewidth}
  \centering
  \centerline{\includegraphics[width=\linewidth]{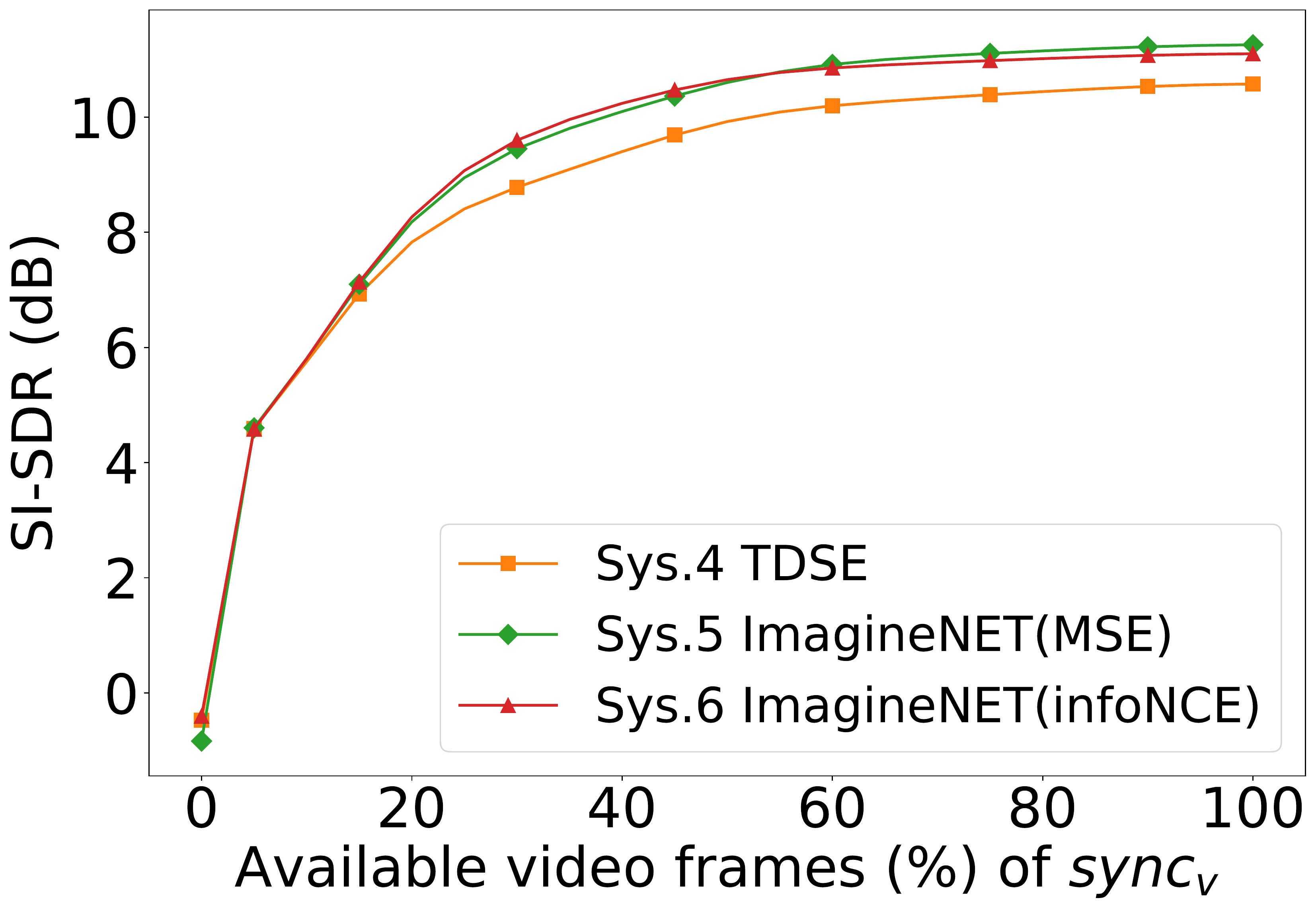}}
  \vspace*{-2mm}
  \caption{\footnotesize The SI-SDR for different percentages of available video frames when $sync_v$ embedding is used.}\medskip
  \label{fig:occl_syncv}
\end{minipage}
\vspace*{-4mm}
\end{figure}

\vspace{-.15cm}
\section{Results}
\label{sec:results}
\vspace{-.1cm}

We use the SI-SDR and signal-to-distortion ratio (SDR) to measure the signal quality of the extracted speech. We also use the perceptual evaluation of speech quality (PESQ) and the short term objective intelligibility (STOI) to evaluate the perceptual quality and intelligibility of the extracted speech. The higher the better for all metrics.

\vspace{-.2cm}
\subsection{Comparison with baseline}
\vspace{-.1cm}

In Table~\ref{tab:baseline}, we compare our ImagineNET with the baseline TDSE network. When $vsr_v$ is used, ImagineNET (Sys. 2 \& 3) outperforms TDSE (Sys. 1) for all evaluation metrics. When $sync_v$ is used, ImagineNET (Sys. 5 \& 6) outperforms TDSE (Sys. 4) for all evaluation metrics. The ImagineNET trained with InfoNCE loss performs similarly to ImagineNET which is trained with MSE loss for both $vsr_v$ and $sync_v$ embeddings. 

In Fig.~\ref{fig:occl_vsr} and~\ref{fig:occl_syncv}, we plot the SI-SDR for different percentages of available video frames when $vsr_v$ and $sync_v$ are used respectively.
The plotted average SI-SDR values are for each 5\% interval of available video frames. More precisely, the first point is the average performance for 0-5\%, and the last point is the average performance for 96-100\%. In Fig.~\ref{fig:occl_vsr} when $vsr_v$ is used, TDSE performs better when only a few video frames are available, but once the percentage of available video frames exceeds 10\%, ImagineNET outperforms TDSE. In Fig.~\ref{fig:occl_syncv} when $synv_v$ is used, all systems perform similarly when the amount of available video frames is small, but ImagineNET outperforms the TDSE once the portion of available video frames exceeds 20 \%. 

The results in Table 1 as well as Fig. 2 and 3 show that the VSR embedding $vsr_v$ outperforms the visual-only SLSyn embedding $synv_v$. It is worth noting that the \textit{reentry} model originally proposed to use the audio-visual SLSyn embedding which outperformed the use of $vsr_v$. However, inpainting audio-visual embeddings is challenging due to the volatile nature of speech signals and, thus, out of the scope of this paper.


\begin{table}
    \centering
    \vspace*{-3mm}
    \caption{Ablation studies for our ImagineNET (with MSE as V loss and $vsr_v$ embedding). We study the contribution of the V loss by varying $\gamma$ in Eq. (\ref{eqa:overall}). We also study the visual inpainting with (\cmark) and without (\xmark) using the estimated intermediate speech in the visual refiners (Use ES).}
    \addtolength{\tabcolsep}{-3.5pt}
    \resizebox{0.95\linewidth}{!}{
    \begin{tabularx}{\linewidth}{Y|Y|l|l|Y|Y|Y} 
       \toprule
        Sys.  &$\gamma$  &\mbox{Use ES} &\mbox{SI-SDR} &SDR  &PESQ   &STOI\\
        \midrule
        2   &1  & \multicolumn{1}{c|}{\cmark} & \multicolumn{1}{c|}{10.81} &11.33   &2.894  &0.854 \\ 
        \midrule
        10  &0  & \multicolumn{1}{c|}{\cmark} & \multicolumn{1}{c|}{10.34}	&10.85	&2.830	&0.846 \\ 

        \midrule
        11  &1  & \multicolumn{1}{c|}{\xmark} & \multicolumn{1}{c|}{\ \ 9.70}	&10.21	&2.764	&0.834 \\
       \bottomrule
    \end{tabularx}
    }
    \addtolength{\tabcolsep}{3.5pt}
    \vspace*{-3mm}
    \label{tab:ablation}
\end{table}

\vspace{-.2cm}
\subsection{Ablation studies}
\vspace{-.1cm}

In Table~\ref{tab:baseline}, we compare our ImagineNET with the baseline TDSE when there are no missing video frames during training and inference. The results show that ImagineNET (Sys. 8 \& 9) outperforms TDSE (Sys. 7) most of the time, although the improvements are not as high as under conditions with missing visual frames (Sys. 1-6).

We also study the contribution of the visual inpainting loss in Table 2. Sys. 10 is equal to Sys. 2 except that we set $\gamma$ to zero, which is equivalent to not using any visual inpainting loss. The results show a drop in performance, showing the importance of the visual inpainting loss. In Sys. 11, we do not feed the estimated intermediate speech into every mask refiner in order to study the contribution of the interaction between the two tasks. The performance drops, showing the importance of the interaction between the two tasks.

\vspace{-.15cm}
\section{Conclusion}
\label{sec:conclusion}
\vspace{-.1cm}

We investigated the time-domain speaker extraction algorithm with intermittent visual cue. We proposed to jointly optimize speaker extraction and the visual embedding inpainting to explore the interaction between the two tasks with an interlacing structure. The results on viseme-phoneme mapping and speech-lip synchronization embeddings with both MSE and infoNCE inpainting loss show that our ImagineNET outperforms the baseline TDSE model. This confirms the design of our method.




\footnotesize
\bibliographystyle{IEEEbib}
\bibliography{IEEEabrv,Bibliography}

\begin{thebibliography}{10}
\providecommand{\url}[1]{#1}
\csname url@rmstyle\endcsname
\providecommand{\newblock}{\relax}
\providecommand{\bibinfo}[2]{#2}
\providecommand\BIBentrySTDinterwordspacing{\spaceskip=0pt\relax}
\providecommand\BIBentryALTinterwordstretchfactor{4}
\providecommand\BIBentryALTinterwordspacing{\spaceskip=\fontdimen2\font plus
\BIBentryALTinterwordstretchfactor\fontdimen3\font minus
  \fontdimen4\font\relax}
\providecommand\BIBforeignlanguage[2]{{%
\expandafter\ifx\csname l@#1\endcsname\relax
\typeout{** WARNING: IEEEtran.bst: No hyphenation pattern has been}%
\typeout{** loaded for the language `#1'. Using the pattern for}%
\typeout{** the default language instead.}%
\else
\language=\csname l@#1\endcsname
\fi
#2}}

\bibitem{tao2021someone}
R.~Tao, Z.~Pan, R.~K. Das, X.~Qian, M.~Z. Shou, and H.~Li, ``Is someone
  speaking? {E}xploring long-term temporal features for audio-visual active
  speaker detection,'' in \emph{Proc. of the 29th ACM Int. Conf. on
  Multimedia}, 2021, pp. 3927--3935.

\bibitem{qian2021multi}
X.~Qian, M.~Madhavi, Z.~Pan, J.~Wang, and H.~Li, ``Multi-target {DoA}
  estimation with an audio-visual fusion mechanism,'' in \emph{Proc. IEEE Int.
  Conf. Acoust., Speech, Signal Process.}, 2021, pp. 4280--4284.

\bibitem{wang2022predict}
J.~Wang, X.~Qian, and H.~Li, ``Predict-and-update network: Audio-visual speech
  recognition inspired by human speech perception,'' \emph{arXiv preprint
  arXiv:2209.01768}, 2022.

\bibitem{gao2021tran}
X.~Gao, C.~Gupta, and H.~Li, ``Automatic lyrics transcription of polyphonic
  music with lyrics-chord multi-task learning,'' \emph{IEEE/ACM Trans. Audio,
  Speech, Lang. Process.}, vol.~30, pp. 2280--2294, 2022.

\bibitem{pan2020multi}
Z.~Pan, Z.~Luo, J.~Yang, and H.~Li, ``Multi-modal attention for speech emotion
  recognition,'' in \emph{Proc. INTERSPEECH}, 2020, pp. 364--368.

\bibitem{luo2019conv}
Y.~{Luo} and N.~{Mesgarani}, ``Conv-{TasNet}: Surpassing ideal time–frequency
  magnitude masking for speech separation,'' \emph{IEEE/ACM Trans. Audio,
  Speech, Lang. Process.}, vol.~27, no.~8, pp. 1256--1266, 2019.

\bibitem{hershey2016deep}
J.~R. {Hershey}, Z.~{Chen}, J.~{Le Roux}, and S.~{Watanabe}, ``Deep clustering:
  Discriminative embeddings for segmentation and separation,'' in \emph{Proc.
  IEEE Int. Conf. Acoust., Speech, Signal Process.}, 2016, pp. 31--35.

\bibitem{von2022sasdr}
T.~von Neumann, K.~Kinoshita, C.~Boeddeker, M.~Delcroix, and R.~Haeb-Umbach,
  ``{SA-SDR}: A novel loss function for separation of meeting style data,'' in
  \emph{Proc. IEEE Int. Conf. Acoust., Speech, Signal Process.}, 2022, pp.
  6022--6026.

\bibitem{wang2023time}
T.~Wang, Z.~Pan, M.~Ge, Z.~Yang, and H.~Li, ``Time-domain speech separation
  networks with graph encoding auxiliary,'' \emph{IEEE Signal Processing
  Letters}, 2023.

\bibitem{wang2019voicefilter}
Q.~Wang, H.~Muckenhirn, K.~Wilson, P.~Sridhar, Z.~Wu, J.~R. Hershey, R.~A.
  Saurous, R.~J. Weiss, Y.~Jia, and I.~L. Moreno, ``{VoiceFilter}: Targeted
  voice separation by speaker-conditioned spectrogram masking,'' in \emph{Proc.
  INTERSPEECH}, 2019, pp. 2728--2732.

\bibitem{vzmolikova2019speakerbeam}
K.~{Žmolíková}, M.~{Delcroix}, K.~{Kinoshita}, T.~{Ochiai}, T.~{Nakatani},
  L.~{Burget}, and J.~{Černocký}, ``Speaker{B}eam: Speaker aware neural
  network for target speaker extraction in speech mixtures,'' \emph{IEEE
  Journal of Selected Topics in Signal Processing}, vol.~13, no.~4, pp.
  800--814, 2019.

\bibitem{Chenglin2020spex}
C.~{Xu}, W.~{Rao}, E.~S. {Chng}, and H.~{Li}, ``Sp{E}x: Multi-scale time domain
  speaker extraction network,'' \emph{IEEE/ACM Trans. Audio, Speech, Lang.
  Process.}, vol.~28, pp. 1370--1384, 2020.

\bibitem{ephrat2018looking}
A.~Ephrat, I.~Mosseri, O.~Lang, T.~Dekel, K.~Wilson, A.~Hassidim, W.~T.
  Freeman, and M.~Rubinstein, ``Looking to listen at the cocktail party: a
  speaker-independent audio-visual model for speech separation,'' \emph{ACM
  Transactions on Graphics}, vol.~37, no.~4, pp. 1--11, 2018.

\bibitem{wu2019time}
J.~{Wu}, Y.~{Xu}, S.~{Zhang}, L.~{Chen}, M.~{Yu}, L.~{Xie}, and D.~{Yu}, ``Time
  domain audio visual speech separation,'' in \emph{Proc. IEEE Autom. Speech
  Recognit. Understanding Workshop}, 2019, pp. 667--673.

\bibitem{pan2021reentry}
Z.~Pan, R.~Tao, C.~Xu, and H.~Li, ``Selective listening by synchronizing speech
  with lips,'' \emph{IEEE/ACM Trans. Audio, Speech, Lang. Process.}, vol.~30,
  pp. 1650--1664, 2022.

\bibitem{usev21}
------, ``{USEV}: Universal speaker extraction with visual cue,''
  \emph{IEEE/ACM Trans. Audio, Speech, Lang. Process.}, vol.~30, pp.
  3032--3045, 2022.

\bibitem{afouras2018conversation}
T.~Afouras, J.~S. Chung, and A.~Zisserman, ``The conversation: Deep
  audio-visual speech enhancement,'' in \emph{Proc. INTERSPEECH}, 2018, pp.
  3244--3248.

\bibitem{Li2020}
C.~Li and Y.~Qian, ``Listen, watch and understand at the cocktail party:
  Audio-visual-contextual speech separation,'' in \emph{Proc. INTERSPEECH},
  2020, pp. 1426--1430.

\bibitem{tavcse2022}
J.~Li, M.~Ge, Z.~Pan, L.~Wang, and J.~Dang, ``{VCSE}: Time-domain
  visual-contextual speaker extraction network,'' in \emph{Proc. INTERSPEECH},
  2022, pp. 906--910.

\bibitem{pan2020muse}
Z.~Pan, R.~Tao, C.~Xu, and H.~Li, ``Mu{SE}: Multi-modal target speaker
  extraction with visual cues,'' in \emph{Proc. IEEE Int. Conf. Acoust.,
  Speech, Signal Process.}, 2021, pp. 6678--6682.

\bibitem{marvin2021}
M.~Borsdorf, C.~Xu, H.~Li, and T.~Schultz, ``Universal speaker extraction in
  the presence and absence of target speakers for speech of one and two
  talkers,'' in \emph{Proc. INTERSPEECH}, 2021.

\bibitem{pan2022hybrid}
Z.~Pan, M.~Ge, and H.~Li, ``A hybrid continuity loss to reduce over-suppression
  for time-domain target speaker extraction,'' in \emph{Proc. INTERSPEECH},
  2022, pp. 1786--1790.

\bibitem{9053730}
M.~Sadeghi and X.~Alameda-Pineda, ``Robust unsupervised audio-visual speech
  enhancement using a mixture of variational autoencoders,'' in \emph{Proc.
  IEEE Int. Conf. Acoust., Speech, Signal Process.}, 2020, pp. 7534--7538.

\bibitem{9414097}
------, ``Switching variational auto-encoders for noise-agnostic audio-visual
  speech enhancement,'' in \emph{Proc. IEEE Int. Conf. Acoust., Speech, Signal
  Process.}, 2021, pp. 6663--6667.

\bibitem{afouras2019my}
T.~Afouras, J.~S. Chung, and A.~Zisserman, ``My lips are concealed:
  Audio-visual speech enhancement through obstructions,'' in \emph{Proc.
  INTERSPEECH}, 2019, pp. 4295--4299.

\bibitem{wu2022occl}
Y.~Wu, C.~Li, J.~Bai, Z.~Wu, and Y.~Qian, ``Time-domain audio-visual speech
  separation on low quality videos,'' in \emph{Proc. IEEE Int. Conf. Acoust.,
  Speech, Signal Process.}, 2022, pp. 256--260.

\bibitem{pan2022seg}
Z.~Pan, X.~Qian, and H.~Li, ``Speaker extraction with co-speech gestures cue,''
  \emph{IEEE Signal Processing Letters}, vol.~29, pp. 1467--1471, 2022.

\bibitem{kim2022visualimpainting}
K.~Kim, J.~Jung, W.~J. Kim, and S.-E. Yoon, ``Deep video inpainting guided by
  audio-visual self-supervision,'' in \emph{Proc. IEEE Int. Conf. Acoust.,
  Speech, Signal Process.}, 2022, pp. 1970--1974.

\bibitem{liu2021fuseformer}
R.~Liu, H.~Deng, Y.~Huang, X.~Shi, L.~Lu, W.~Sun, X.~Wang, J.~Dai, and H.~Li,
  ``Fuseformer: Fusing fine-grained information in transformers for video
  inpainting,'' in \emph{Proc. ICCV}, 2021, pp. 14\,040--14\,049.

\bibitem{oord2018representation}
A.~v.~d. Oord, Y.~Li, and O.~Vinyals, ``Representation learning with
  contrastive predictive coding,'' \emph{arXiv preprint arXiv:1807.03748},
  2018.

\bibitem{le2019sdr}
J.~Le~Roux, S.~Wisdom, H.~Erdogan, and J.~R. Hershey, ``{SDR}--half-baked or
  well done?'' in \emph{Proc. IEEE Int. Conf. Acoust., Speech, Signal
  Process.}, 2019, pp. 626--630.

\bibitem{morrone2021audio}
G.~Morrone, D.~Michelsanti, Z.-H. Tan, and J.~Jensen, ``Audio-visual speech
  inpainting with deep learning,'' in \emph{Proc. IEEE Int. Conf. Acoust.,
  Speech, Signal Process.}, 2021, pp. 6653--6657.

\bibitem{8867915}
A.~Marafioti, N.~Perraudin, N.~Holighaus, and P.~Majdak, ``A context encoder
  for audio inpainting,'' \emph{IEEE/ACM Trans. Audio, Speech, Lang. Process.},
  vol.~27, no.~12, pp. 2362--2372, 2019.

\bibitem{kingma2013auto}
D.~P. Kingma and M.~Welling, ``Auto-encoding variational bayes,'' \emph{arXiv
  preprint arXiv:1312.6114}, 2013.

\bibitem{elharrouss2020image}
O.~Elharrouss, N.~Almaadeed, S.~Al-Maadeed, and Y.~Akbari, ``Image inpainting:
  A review,'' \emph{Neural Processing Letters}, vol.~51, no.~2, pp. 2007--2028,
  2020.

\bibitem{he2020momentum}
K.~He, H.~Fan, Y.~Wu, S.~Xie, and R.~Girshick, ``Momentum contrast for
  unsupervised visual representation learning,'' in \emph{Proc. CVPR}, 2020,
  pp. 9729--9738.

\bibitem{Chung18b}
J.~S. Chung, A.~Nagrani, and A.~Zisserman, ``{VoxCeleb2}: Deep speaker
  recognition,'' in \emph{Proc. INTERSPEECH}, 2018, pp. 1086--1090.

\end{thebibliography}

\end{document}